\documentstyle[12pt]{article}
\begin{document}
\baselineskip=24pt
\rightline{\bf Preprint No. imsc-92/}
\vskip 5em
\begin{center}
\LARGE{\bf BOSONISATION IN ANY DIMENSION}
\vskip 5em
\Large{\bf H.S.SHARATCHANDRA}
\vskip 2em
\large{\bf The Institute of Mathematical Sciences, \\
Madras-600 113, INDIA}
\end{center}
\vskip 15em
\leftline{\bf E-mail address: sharat@imsc.ernet.in}
\newpage

ABSTRACT

The mechanism underlying any bosonisation or fermionisation is
exposed.It is shown that any local theory of fermions on a lattice
in any spatial dimension greater than one is equivalent to a
local theory of Ising spins coupled to a $Z_{2}$ gauge field.There
is a close relation to the descrription of anyons using a
Chern-Simmons term.

\newpage

It has been discovered time and again that some models
with fermions in quantum field theory and statistical
mechanics are equivalent to some other models with only
bosons as the statistical or dynamical variables.Also
there are models with only bosonic variables having fermions
in the spectrum.These two phenomena may be termed
'bosonisation'  and 'fermionisation'  respectively.Some of the
classic
examples are   as follows.Two- dimensional Ising
model can be rewritten \cite{is2}  (and solved)  us\-ing
fermionic
var\-iables.The Thirring and the Schwinger models of fermions in 1+1
-dimensions  are solved \cite{thsc} using free bosonic fields.
The Luttinger model of interacting fermions in 1+1 -dimensions
can be rewritten  \cite{lutt} as a certain theory of bosons.
The  sine-Gordon
model in 1+1- dimensions is exactly equivalent \cite{sg} to the
massive Thirring model. In 2+1-dimensions non-linear $\sigma$ -
models with Hopf term \cite{wz} of a specific strength has fermionic
excitations.Anyons described by using a 2+1-dimensional
 Chern-Simmons field
theory with only bosonic variables \cite{any}  is
actually a theory of fermions for a specific value of the
statistical parameter.Bound states of spin zero magnetic
monopoles and electric
charges may carry half integral spin \cite{mon}.As a consequence
certain
non-abelian gauge theories in 3+1- dimensions have fermions in
the spectrum \cite{non} .Non-linear $\sigma$ - models in 3+1- dimensions
may have fermionic excitations described semi-classically
by the Skyrmion solutions \cite{sky} .

The phenomenon of bosonisation or fermionisation has always
appeared mysterious and context bound.I take out some of
this mystery here.The equivalence is already at the kinematic
level and is simply a consequence of the Jordan-Wigner transformation
\cite{jw}.I show that any local theory of fermions in any spatial
dimension greater than one is equivalent to a local theory of bosons
coupled to an Abelian gauge field.  I will use the barest
 essentials and
demonstrate this for Hamiltonians on a lattice
 with one species of fermions
in two-  and three- dimensions.I present a simple analysis, closely
following the arguments used for 2-dimensional Ising model.

There have been many different attempts with various motivations
at bosonisation and fermionisation in 2+1- and 3+1- dimensions.See
Ref \cite{misc} for some of these attempts.My method has some points in
common with, but my results are
different from all these and emphasizes the generality of
the phenomenon independent of dimension and the crucial role
played by the Abelean gauge field.

I begin with the 2-dimensional Ising model in the Hamiltonian
formalism.The Hamiltonian is,
\begin{equation}
H=\sum_{\bf n}( \lambda \sigma_{1}({\bf n}) \sigma_{1}({\bf n}) +
\sigma_{3}({\bf n}))
  \label{NU1}
\end{equation}
The integers {\bf n} label the sites on a line.Variables
$\sigma_{a}({\bf n})$ , a=1,2,3
 at different sites commute with each other,
\begin{equation}
\sigma_{a}({\bf m})  \sigma_{b} ({\bf n}) =
\sigma_{b}({\bf n})  \sigma_{a}({\bf m}) , {\bf m} \neq {\bf n}
\end{equation}
whereas at a given site,
\begin{equation}
\sigma_{1}({\bf n})  \sigma_{2}({\bf n}) =
- \sigma_{2}({\bf n}) \sigma_{1}({\bf n}) = i \sigma_{3}({\bf n})
\end{equation}
 and other equations obtained by a cyclic interchange of the
subscripts.
The first term on the right hand side of  Eq.~\ref{NU1} is the 'potential'
energy and the second term is the 'kinetic energy'.I have deliberately
interchanged $\sigma_{1}$ and $\sigma_{3}$ in the standard form for later
convenience.
Define raising and lowering operators,
\begin{equation}
\sigma_{\pm}({\bf n}) = \frac{1}{2}(\sigma_{1}({\bf n})
 \pm \sigma_{2}({\bf n}))
\end{equation}
and a state $\mid o >$ by,
\begin{equation}
\sigma_{-}({\bf n}) \mid o >  = 0 .
\end{equation}
This state $\mid o>$ has spin pointing 'down' at every  site.
A set of basis vectors  for the Hilbert space is obtained by
upturning some of the spins:
\begin{equation}
\sigma_{+}({\bf m}) \sigma_{+}({\bf n}) \cdots \sigma_{+}({\bf r})
 \mid o >
\end{equation}
Notice that this Hilbert state is in 1:1
 correspondence with
that for a theory of fermions defined by associating
 an annihilation
operator $\psi_{-}({\bf n})$ and a creation operator
$\psi_{+}({\bf n})$ at each site {\bf n}:
\begin{equation}
\sigma_{\pm}({\bf n})  \rightarrow \psi_{\pm}({\bf n}) \label{2}
\end{equation}
The state with all spins down is mapped to the vacuum
$\mid O>$ of
the fermionic theory.The state with up spins at sites
$ {\bf m},{\bf n}, \cdots ,{\bf r}$
is mapped on to the state with fermions occupying the
 corresponding
sites.

However the analogy is not complete at this stage.We have,
\begin{eqnarray}
\{  \psi_{-}({\bf m}), \psi_{+}({\bf n}) \} & = &
 \delta_{{\bf m}, {\bf n}}    \nonumber  \\
\{ \psi_{-}({\bf m}),  \psi_{-}({\bf n}) \} & = & 0 ,\nonumber
\\
\{ \psi_{+}({\bf m}),  \psi_{+}({\bf n}) \} & = & 0
\end{eqnarray}
On the other hand even though the corresponding $\sigma$ -variables
at the same site have this algebra, they commute with each other
at different sites in contrast to the the anti-commuting $\psi$'s.

The 1:1 mapping of the basis states is still correct with a choice of
ordering the fermion operators. For example , we may 'normal order'
the fermions $\psi_{+}({\bf n})$ such that $\psi_{+}({\bf n})$
 is to the left of $\psi_{+}({\bf m})$ if ${\bf n} <  {\bf m}$.
However the operators related by the mapping Eq.~\ref{2}
 are different.
The matrix elements in the basis states differ by a sign at times,
because of the need to normal order.
It is well known \cite{jw}  that this difficulty can be overcome
 in the present
case.Using a chain of $\sigma_{3}({\bf p})$ 's the
 mapping can be made exact:
\begin{equation}
\psi_{\pm}({\bf n})  =( \prod_{{\bf p}<{\bf n}}
 \sigma_{3}({\bf p}) )  \sigma_{\pm}({\bf n})  \label{3}
\end{equation}
Consider
 $\psi_{\pm}({\bf m}) \psi_{\pm}({\bf n}),{\bf m} < {\bf n}$
In order  to move $\psi_{\pm}({\bf n})$  to the left of
$\psi_{\pm}({\bf m})$ , we have to move $\sigma_{3}({\bf n})$ to
 the left of $\sigma_{\pm}({\bf m})$.This gives
anti-commutativity.All anti-commutation relations
 are reproduced by
this mapping.Also,
\begin{eqnarray}
\sigma_{1}({\bf n})  \sigma_{1}({\bf n+1})  &  & \nonumber  \\
&  = & \frac {1}{2} ( \sigma_{+}({\bf n}) +
 \sigma_{-}({\bf n}))
 \sigma_{3}({\bf n+1})
 \frac{1}{2}
 (\sigma_{+}({\bf n+1}) -  \sigma_{-}({\bf n+1}))
 \nonumber  \\
& = & \frac{1}{4} (\psi_{+}({\bf n}) + \psi_{-}({\bf n}))
 (\psi_{+}({\bf n+1}) - \psi_{-}({\bf n+1}))  \\
\frac{1}{2}(1 \pm \sigma_{3}({\bf n})) & = &
\sigma_{\pm}({\bf n}) \sigma_{\mp}({\bf n})  \nonumber \\
 & =  &  \psi_{\pm}({\bf n}) \psi_{\mp}({\bf n})
\end{eqnarray}
Thus ,
\begin{equation}
H  =  \sum_{\bf n}(\frac{\lambda}{4} (\psi_{+}({\bf n})
+ \psi_{-}({\bf n}))(\psi_{+}({\bf n+1})
-  \psi_{-}({\bf n+1})) +
[\psi_{+}({\bf n}), \psi_{-}({\bf n})])
\end{equation}
This way the 2-dimensional Ising model is equi\-va\-lent
to a lo\-cal
th\-eory of fermions.

I show that the same arguments can be used
 in higher dimensions
with one additional input.Consider one species of fermions
$\psi_{\pm}({\bf n})$ on a two dimensional
finite square lattice with free boundary conditions.
Exactly as before ,
there is a 1:1 mapping between the Fock state basis
 of this theory
and the configurations of  an Ising model on the
 same lattice
once a normal ordering is chosen for the fermions.A
 convenient
choice for us is as follows.Join all lattice sites as shown
by the thick lines in the Figure.This way every site is
covered and each site is visited once only, i.e. we have a
maximal tree.Define ${\bf m} < {\bf n}$  if the
 site {\bf m} is visited before the  site {\bf n}.
The canonical basis is then ,
\begin{equation}
\psi_{+}({\bf m}) \psi_{+}({\bf n}) \cdots
  \psi_{+}({\bf r}) \mid O>
\end{equation}
where ${\bf m}<{\bf n}< \cdots <{\bf r}$ .As before the
 mapping Eq.~\ref{3}
 gives a
bosonisation.Now ${\bf p}<{\bf n}$ refers to the
ordering defined above.

Now comes a point of departure from 1+1 -dimensional case.
Consider a fermion Hamiltonian such as ,
\begin{eqnarray}
H &=  & \sum_{\bf n}  ( \psi_{+}({\bf n})  \psi_{-}({\bf n})
 + \kappa  \sum_{{\bf n},{\bf i}}
( \psi_{+}({\bf n}) \psi_{-}({\bf n+i}) + h.c. ) \label{hfer}
\end{eqnarray}
where ${\bf i} = {\bf 1},{\bf 2}$ denotes the unit vectors
in $\bf 1$- and $\bf 2$- directions.
We want the corresponding Hamiltonian in the Ising variables.
As before ,
\begin{eqnarray}
\psi_{+}({\bf n}) \psi_{-}({\bf n+1})
& \longrightarrow & \sigma_{-}({\bf n})  \sigma_{-}({\bf n+1})
\  if\ n_{2}\ is\ odd
\nonumber \\
& \longrightarrow  &  \sigma_{+}({\bf n}) \sigma_{+}({\bf n})
 \ if\ n_{2}\ is\ even
\end{eqnarray}
But the other 'hopping term',
 $\psi_{+}({\bf n}) \psi_{-}({\bf n+2})$
looks non-local in the Ising variables.  $\sigma_{3}$
 variables on on all sites of the
tree from ${\bf n}$ to ${\bf n+2}$ appear.With our
 normal ordering, neighboring
variables in the $\bf 2$-direction have become far distant from each
other.

At this stage it appears that though bosonisation is possible
a local Hamiltonian gets mapped into a non-local
 Hamiltonian.We
now show that this problem can be re\-moved 
us\-ing a $Z_{2}$ gauge field \cite{z2}.
In\-tro\-duce $\mu_{3}({\bf n},{\bf i}) \equiv
  \mu_{3}({\bf n+i},{\bf -i}),
{\bf i=1,2}$, living on the links of the lattice
and taking values $\pm 1$.The local gauge transformation is,
\begin{eqnarray}
\mu_{3}({\bf n},{\bf i}) &  \longrightarrow &   \nu({\bf n})
 \mu_{3}({\bf n},{\bf i})
  \nu({\bf n+i}), \nonumber  \\
\sigma_{\pm}({\bf n}) &  \longrightarrow &    \nu({\bf n})
 \sigma_{\pm}({\bf n})
\end{eqnarray}
where  $\nu({\bf n}) =\pm 1$ is the local gauge parameter.I
 show that the fermion Hamiltonian Eq.~\ref{hfer}
is equivalent to the {\em local} Hamiltonian
\begin{equation}
H  = \sum_{\bf n} \frac{1}{2}(1+\sigma_{3}({\bf n}))+\kappa
\sum_{{\bf n},{\bf i}} ( \sigma_{+}({\bf n})
 \mu_{3}({\bf n},{\bf i}) \sigma_{+}({\bf n+i})+ h.c.)
\label{hbos}
\end{equation}
with the {\em local} constraint,
\begin{eqnarray}
\prod_{P({\bf n})} \mu_{3}({\bf n},{\bf i}) & = &
 \sigma_{3}({\bf n})  \sigma_{3}({\bf n+2})
 \ if\ n_{2}\ is\ odd,  \nonumber  \\
& & =\sigma_{3}({\bf n+1}) \sigma_{3}({\bf n+1+2})
 \ if\ n_{2}\ is\ even  \label{con}
\end{eqnarray}
where $P({\bf n})$ is the plaquette formed by the vertices,
({\bf n},{\bf n+1},{\bf n+2},{\bf n+1+2}).

There is an additional clarification. The constraint
has to be modified for some of the edge plaquettes.They
are those which border the maximal tree as it crosses from
one row to the next.
(They are marked with  the  very thick lines in the
Figure.) For such plaquettes the r.h.s.\ of the constraint
equation Eq.~\ref{con} is the product of all four spins
at its vertices.

The proof of this equivalence is very simple.We gauge fix
all link variables on our maximal tree (of the new lattice)
to +1.Then $\mu_{3}({\bf n},{\bf 1}) =+1$
everywhere.Further $\mu_{3}({\bf n},{\bf 2})$ can be easily
 calculated in terms of
$\sigma_{3}$'s by simply multiplying the row of plaquettes
 to the right
(left) of the link  if $n_{2}$ is odd (even) and noting
 that the last
vertical link in this product is gauge fixed to +1. We get
precisely the chain of $\sigma_{3}({\bf p})$'s on the
 tree connecting
 $\bf n$ to $\bf n+2$.
Thus we have reproduced all terms coming from the replacement
Eq.~\ref{3} correctly.

Once we have shown the equivalence in one gauge, we may expect
 the same in
any other gauge or even without any gauge fixing.But now
 there is an
additional problem.It appears that the constraint Eq.~\ref{con}
does not commute
with the the Hamiltonian Eq.~\ref{hbos}.This is because
 $\sigma_{\pm}({\bf n})$ doesnot
commute with the right hand side of Eq.~\ref{con} for
 an appropriate
{\bf n} whereas $\mu_{3}({\bf n},{\bf i})$'s are presumed to
commute with
each other.This serious problem can be overcome by postulating
that $\mu_{3}({\bf n},{\bf i})$'s  for various links donot
 all commute
 with each other.To
be specific, (see Figure),
\begin{eqnarray}
\{\mu_{3}({\bf n},\pm {\bf 2}), \mu_{3}({\bf n},\pm {\bf 1})\} = 0
\  if\ n_{2}\ is\ even, \nonumber \\
\{\mu_{3}({\bf n},\pm {\bf 2}), \mu_{3}({\bf n},\mp {\bf 1})\} = 0
\  if\ n_{2}\ is\ odd
\end{eqnarray}
This means $\mu_{3}({\bf m},{\bf 2})$'s and
 $\mu_{3}({\bf n},{\bf 1})$'s
 are more like conjugate variables
instead of being
independent variables.A way of handling this situation is as
follows.Define the triplet
 $\mu_{a}({\bf n},{\bf 1})$
 a=1,2,3 as
satisfying Pauli algebra for each $({\bf n},{\bf i})$.Then our
 requirements can be
satisfied by the  replacement,
\begin{eqnarray}
\mu_{3}({\bf n},{\bf 2})
& \longrightarrow &
  \mu_{1}({\bf n},{\bf 1})
  \mu_{3}({\bf n},{\bf 2}) \mu_{1}({\bf n+2},{\bf 1})
 \  if\ n_{2}\ is\ even, \nonumber  \\
&    \longrightarrow  &  \mu_{1}({\bf n},{\bf -1})
  \mu_{3}({\bf n},{\bf 2}) \mu_{1}({\bf n+2},{\bf -1})
\  if\ n_{2}\ is\ odd.
\end{eqnarray}
where now $\mu_{3}({\bf n},{\bf i})$ 's are presumed to
commute with each other.
Now we have to be more careful in ordering the terms
 in Eq.~\ref{con}
because $\mu_{3}({\bf m},{\bf 1})$'s may not commute with
 $\mu_{3}({\bf n},{\bf 2})$'s.
It is suffi\-cient to put
the two $\mu_{3}({\bf n},{\bf 1})$ 's together.
$\mu_{3}({\bf n},{\bf 2})$'s
  which always commmutes with this pair ,
may be placed on either side of the pair.

That $\mu_{3}({\bf m},{\bf 1})$'s and
 $\mu_{3}({\bf n},{\bf 2})$'s
  may not commute has close analogy with the
commutation relations,
\begin{equation}
[A_{1}({\bf x}),A_{2}({\bf y})]=i \delta ^{2}({\bf x}-{\bf y})
\end{equation}
in anyon dynamics.Moreover the constraint Eq.~\ref{con}
 is the  analogue of
the constraint,
\begin{equation}
F_{12}({\bf x}) =  j_{0}({\bf x})  \label{any}
\end{equation}
This is not just a coincidence. The underlying mechanism is
essentially the same.Note that the constraint Eq.~\ref{any} is a
consequence of the Chern-Simons term in the action.Infact
the zeroeth component of vector potential acts as the Lagrange
multiplier for the the constraint.In the same way the constraint
Eq.~\ref{any} can be obtained from a $Z_{2}$ link variable on
 time-like links.
In the continuum formulation there is a scalar field in
place of the Ising variables and the $Z_{2}$ gauge invariance
gets promoted to an U(1) gauge invariance.

The fermi field may be expressed as the gauge invariant
object,
\begin{equation}
\psi_{\pm}({\bf n}) = (\prod_{tree{\bf n}}
\mu_{3}({\bf n},{\bf i}))\sigma_{\pm} ({\bf n})
\end{equation}
where the product is taken along the maximal tree upto $\bf n$.
This is simply the Ising variable dressed with the 'Coulomb'
field resulting from the constraint equation, Eq.~\ref{con}.
This equation is readily verified in the gauge in which all
$\mu_{3}({\bf n},{\bf 2})$ are set to +1.In this gauge,
$\mu_{3}({\bf n},{\bf 1})$ is  equal to $\sigma_{3}$
-variable at one of its vertices.

I now argue that the results can be easily extended to higher
dimensions by considering 3+1- dimensional case in particular.
Again I consider a finite cubic lattice and free boundary
 conditions.
On each 1-2 plane choose a maximal tree as in the two
 dimensional
case.In each plane the end of the tree continues along
the vertical
link and starts the tree in the 1-2 plane above.This tree
provides
the normal ordering for the fermions.Constraints of the type
Eq.~\ref{con} are imposed on the plaquettes in 2-3
and 3-1 planes
also.A
local bosonic Hamiltonian is obtained exactly as in 2+1-
dimensions.

My techniques can be applied to theories with more than one
species of fermions in a straightforward way.We need
 a separate
 Ising spin for each species but a common $Z_{2}$ gauge
field.The r.h.s of the constraint Eq.~\ref{con} has a product of
contributions from each species.My techniques also map any
theory with Ising type variables onto a local theory
of fermions.Note that once the anti-commutation is realized,
obtaining the correct spins for the fermions is  relatively
easy.It simply corresponds to a specific way of coupling
different species of fermions.Also more complicated
local interactions are mapped into local interaction
terms.

It is important to develop bosonisation techniques directly
at the level
of the partion function.This will be very useful for numerical
calculations such as Monte Carlo simulations with fermions.It
is also
interesting to extend the techniques to  continuum theories.
I will address these issues elsewhere.

I have shown here that mapping a local theory of fermions into
a local theory of bosons or vice-versa is  almost as easy
in higher dimensions as in 1+1-dimensions.An abelian gauge
field makes this possible.Its role is similar to that in
electrodynamics which is local inspite of long range Coulomb
interactions.There is a close relation to the description
of anyons using a Chern-Simons term.

\newpage

Figure Caption

The thick line denotes the maximal tree used for ordering the
lattice sites.The extra thick lines at the edges mark the
plaquettes for which the constraint equation needs an edge
correction. A vertical link anti-commutes with the horizontal
links marked by the same letter.

\setlength{\unitlength}{1mm}
\begin{picture}(100,100)(0,0)
\multiput(10,0)(10,0){6}{\line(0,1){50}}
\linethickness{0.5mm}
\multiput(10,0)(0,10){6}{\line(1,0){50}}
\linethickness{1mm}
\multiput(0,0)(0,10){6}{\line(1,0){10}}
\multiput(60,0)(0,10){6}{\line(1,0){10}}
\multiput(0,10)(0,20){2}{\line(0,1){10}}
\multiput(70,0)(0,20){3}{\line(0,1){10}}
\put(10,14){\bf a}
\put(15,11){\bf a}
\put(15,17){\bf a}
\put(20,14){\bf b}
\put(25,11){\bf b}
\put(25,17){\bf b}
\put(44,21){\bf c}
\put(44,27){\bf c}
\put(48,24){\bf c}
\put(54,21){\bf d}
\put(54,27){\bf d}
\put(58,24){\bf d}
\end {picture}
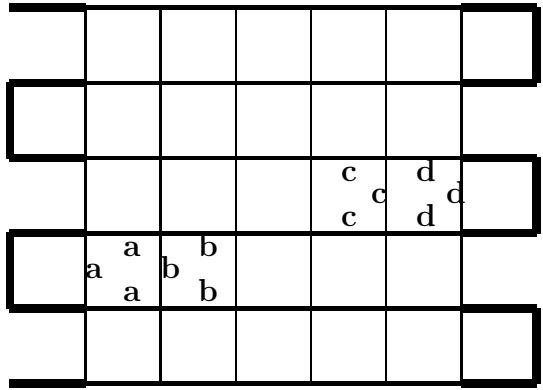
\end{document}